# Study of ion feedback in multi-GEM structures

A. Bondar, A. Buzulutskov [*], L. Shekhtman, A. Vasiljev

*Budker Institute of Nuclear Physics, 630090 Novosibirsk, Russia*

**Abstract**

We study the feedback of positive ions in triple and quadruple Gas Electron Multiplier (GEM) detectors. The effects of GEM hole diameter, detector gain, applied voltages, number of GEMs and other parameters on ion feedback are investigated in detail. In particular, it was found that ion feedback is independent of the gas mixture and the pressure. In the optimized multi-GEM structure, the ion feedback current can be suppressed down to 0.5% of the anode current, at a drift field of 0.1 kV/cm and gain of $10^4$. A simple model of ion feedback in multi-GEM structures is suggested. The results obtained are relevant to the performance of time projection chambers and gas photomultipliers.
*PACs:* 29.40C.

## 1. Introduction

In recent years there have been considerable advances in GEM (Gas Electron Multiplier [1]) detectors in the field of particle tracking [2], photon detection [3,4], radiation imaging [5] and high-pressure detectors [6]. These are based on various physical effects taking place in multi-GEM structures [7].

One such effect is the natural capability of multi-GEM structures to suppress a feedback of positive ions from an avalanche [8,4]. This property would allow to suppress the photocathode degradation in gas photomultipliers (induced by ion impact) [4] and to prevent the field distortion in Time Projection Chambers (TPC) (induced by ion migration into the drift volume) [9].

The latter feature is of primary importance for TPC performance at high particle fluxes, in particular for TESLA TPC [10]. In the ideal case, the ion charge migrated into the drift volume should be of the order of the primary ionization charge, i.e. $F \sim 1/G$, where $F$ is the ion feedback fraction (see next section for its definition) and $G$ is the detector gain.

Ion feedback was studied earlier in single and double GEM structures in $Ar/CO_2$ [8] and in triple and quadruple GEM structures in $Ar/CH_4$

[*] Corresponding author. Fax: 7-3832-342163. Phone: 7-3832-394833.
Email: buzulu@inp.nsk.su

[4]. It was shown that it is mostly sensitive to the detector gain and the drift field. Other parameters, such as the electric field between GEMs, the asymmetry in voltages applied across each GEM and the number of GEM elements had minor effects.

In this paper we further study the ion feedback effects in triple and quadruple GEM structures. We study the effects of the gas mixture, pressure, GEM hole diameter, detector gain and other parameters. Based on the obtained results, the optimized multi-GEM configuration with an enhanced suppression of ion feedback is proposed. We also consider a simple model of ion feedback in GEM detectors. A possible relation of ion feedback reduction observed at high gains to the effect of avalanche extension from the GEM holes is discussed.

## 2. Experimental setup and procedures

Three or four GEM foils and a printed-circuit-board (PCB) anode were mounted in cascade inside a stainless-steel vessel (see Fig.1). The GEMs, produced at CERN workshop, have the following characteristics: 50 μm thick kapton, double conical holes arranged in a hexagonal lattice with a 140 μm pitch, 28×28 mm$^2$ active area. The GEMs of two types were used: with a hole diameter of 85 μm and 40 μm at the metal side and correspondingly 50 μm and 35 μm at the kapton side (at the centre of the hole).

A drift gap (between the cathode and the 1$^{st}$ GEM), transfer gaps (between the GEMs) and an induction gap (between the last GEM and the anode) were 3, 1.6 and 1.6 mm, respectively. The detector was filled with Ar/CH$_4$ (90/10), Ar/CF$_4$ (90/10), Xe/CH$_4$ (90/10) or pure He. In He it could operate at pressures reaching 10 atm; in other gas mixtures the operation pressure was 1 atm.

The detector was irradiated with an X-ray tube through a 1 mm thick Al window. The GEM electrodes were biased through a resistive high-voltage divider, as shown in Fig.1. In He at 1 atm typical electric fields were $E_T \approx 1.1$ kV/cm in the transfer gaps and $E_I \approx 2$ kV/cm in the induction gap; the voltage across a single GEM ("GEM voltage") was $\Delta V_{GEM}$ =150-200 V. In other mixtures the corresponding values were $E_T \approx 2.2$ kV/cm, $E_I \approx 4.4$ kV/cm, $\Delta V_{GEM}$ =300-400 V. The voltage across the drift gap was either equal to that applied across each GEM, $\Delta V_D = \Delta V_{GEM}$, or constant.

The anode signal was recorded in a current mode. The anode current value was always kept below 100 nA, reducing X-ray tube intensity, to prevent charging-up of GEMs. The detector gain was defined as the anode current divided by the current induced by primary ionization in the drift gap. The latter current was determined in special measurements, where the drift gap was operated in an ionization mode.

The ratio of the cathode-to-anode currents provides the ion feedback fraction to the cathode: $F=I_C/I_A$. Here a possible contribution of the primary ionization current to $I_C$ is neglected, since the ion feedback suppression never reached the ultimate limit, i.e. always $F>>1/G$.

It should be remarked that the value of ion feedback fraction in the current work is larger, by a factor of 2-3, compared to that of Refs. [4,8] obtained under similar conditions. In the current work the anode signal was read out from the PCB, while in [4,8] from the bottom electrode of the last GEM. Therefore, in the current work the detected gain is only a fraction (typically 1/3) of the "real" gain due to avalanche charge sharing between the PCB and the bottom face of the last GEM. Following Ref. [4], these modes of operation, for triple GEM detectors, are designated as 3GEM+PCB and 3GEM, respectively.

## 3. Results

In general, the ion feedback might be governed by few parameters (see Fig.1): gas mixture, pressure ($p$), detector gain ($G$), GEM

hole diameter (*d*), drift field ($E_D$), transfer field ($E_T$), number of GEMs and asymmetry in voltages applied across each GEM.

In particular, the electron and ion diffusion, which is generally a function of the gas and the pressure, could affect the charge transfer through a multi-GEM structure. It is known however that the GEM gain rapidly decreases with pressure in gas mixtures with molecular additives [11]. On the other hand, the high-pressure operation of multi-GEM detectors has been recently demonstrated in pure He and Ne [12]. Therefore we used here pure He to study the ion feedback dependence on pressure.

Fig.2 shows gain-voltage characteristics of triple-GEM detectors in He, at 1 and 10 atm, and in $Ar/CH_4$, $Ar/CF_4$ and $Xe/CH_4$, at 1 atm. Two GEM hole configurations were used, with an equal hole diameter, of 85 µm, in all three GEMs and with a reduced hole diameter, of 40 µm, in the middle GEM. Here, the designation "85-40-85 µm" means that the hole diameters are 85, 40 and 85 µm in the $1^{st}$, $2^{nd}$ and $3^{rd}$ GEMs, respectively.

One can see, that for a given voltage the effective (measured) gain turned out to be the same in both configurations. It was also true for other configurations, having reduced holes in other two GEMs. This is obviously due to the effect of saturation of the effective gain observed in [8]: though the real gain increases with decreasing diameter, the effective gain remains saturated, for hole diameters smaller than 70 µm, due to decreasing electron transfer efficiency.

Figs. 3 and 4 illustrate the pressure and gas mixture effects. Fig.3 shows the ion feedback fraction as a function of the gain of a triple GEM detector in He at 1, 5 and 10 atm and in $Ar/CF_4$ at 1 atm. The data were obtained at a drift field proportional to the field inside the GEM hole ($\Delta V_D = \Delta V_{GEM}$). And Fig.4 shows the comparison of ion feedback in $Ar/CH_4$, $Ar/CF_4$ and $Xe/CH_4$ at a constant drift field: $E_D$=0.5 kV/cm.

The results might seem to be unexpected. Indeed, the difference in operation voltages between He and $Ar/CF_4$ is of about a factor of 2 (see Fig.2), while the difference in *E/p* can even reach a factor of 10. Despite of this, the ion feedback is practically the same in all gases and at all pressures, for a given gain. The conclusion is that the electron and ion diffusions, which are functions of the pressure, gas and electric field, do not affect the ion feedback.

The effect of the GEM hole diameter is illustrated in Fig.5, showing the ion feedback fraction as a function of the gain of a triple GEM detector at different GEM hole configurations: 85-85-85 µm, 85-40-85 µm, 40-40-85 µm and 40-40-40 µm. Few statements can be derived.

First, there is a substantial ion feedback decrease when using the middle GEM with reduced holes. However, reducing holes in other GEMs results in increasing the ion feedback, apparently due to increasing the real GEM gain as discussed above.

Second, the ion feedback fraction is with good accuracy an inverse power function of the gain:

$F = bG^{-a}$ ; $a,b > 0$.

This is illustrated by the fact that the data points are fitted well by straight lines in a double-logarithmic scale.

Third, the line slope seems to be defined by the GEM hole diameter: in particular, configurations with reduced holes have similar slopes.

Fourth, at gains higher than $5 \times 10^4$ the given power dependence is violated: the ion feedback abruptly drops down (this is also seen in fig.3 in $Ar/CF_4$).

As we will see in the following section, first three statements may be explained if to adopt the hypothesis that the ion feedback is induced mainly by the middle GEM.

The effect of the drift field is illustrated in Figs.6 and 4. The ion feedback turned out to be much more sensitive to the drift field as compared to other parameters: it increases almost linearly with it. One can see that at a gain of $10^4$ the ion feedback fraction can be suppressed down to 3% and 0.5% at drift fields of 0.5 and 0.1 kV/cm, respectively. These values correspond to the optimal drift fields in Ar/CF$_4$ and Ar/CH$_4$, respectively, at which the electron drift velocity has a maximum.

Fig.7 shows the effect of increasing the number of GEMs. It is interesting that for the triple GEM hole configuration 85-85-85 μm, adding the 4$^{th}$ GEM does not decrease the ion feedback. On the other hand, there is a substantial decrease of ion feedback when adding the GEM to the configuration with reduced holes, 40-40-40 μm. That might mean that the GEM number effect takes place only for the GEMs with reduced charge transfer efficiency.

Fig.8 illustrates the effect of the transfer fields and that of the voltage of the 1$^{st}$ GEM. Here a required voltage configuration is obtained modifying voltage divider shown in Fig.1 ("standard" divider). One can see that the result of varying the transfer field is either negligible, for enhanced transfer fields $E_{T1}$ and $E_{T2}$, or negative, for reduced $E_{T1}$. The noticeable positive effect is obtained only when decreasing the 1$^{st}$ GEM voltage, by 25%.

## 4. Discussion

### 4.1 A model of ion feedback

Let us try to understand the gain dependence of ion feedback in multi-GEM structures using a simplified model. In this model the properties of a GEM element are described by a few parameters (see Fig.9): the ion collection efficiency $\eta$ (the fraction of ions collected into the GEM holes from the gap), the ion extraction efficiency $\varepsilon$ (the fraction of ions extracted from the GEM holes into the gap) and the effective ("visible") GEM gain $g$ as seen by the following GEM or PCB element. The latter parameter is in fact the product of the electron collection efficiency into the holes, the real gain (amplification factor) of the GEM and the electron extraction efficiency from the holes ($\alpha$). We also suppose that in the first approximation $\varepsilon$, $\eta$ and $\alpha$ are independent of the gain, i.e. they are fully defined by the GEM hole geometry and electric fields in the adjacent gaps.

To reduce the number of parameters, we consider here a simplified 3GEM+PCB configuration, often used in practice: with symmetrical double-conical holes and with transfer and induction fields equal to each other, but the drift field being different from them. The appropriate notations are presented in Fig.9.

Then the gain of a triple GEM detector (the number of electrons reaching the PCB) is

(1) $\quad G = N_e = g_1 g_2 g_3 .$

The number of ions backdrifting to the cathode is

(2) $\quad N_i = (g_1/\alpha_1)\varepsilon_1 + (g_1 g_2/\alpha_2)\varepsilon_2\eta_1\varepsilon_1 +$
$+ (g_1 g_2 g_3/\alpha_3)\varepsilon_3\eta_2\varepsilon_2\eta_1\varepsilon_1$

And the ion feedback fraction to the cathode is

(3) $\quad F_{3GEM} = N_i/N_e = \varepsilon_1[\dfrac{1}{\alpha_1 g_2 g_3} + \dfrac{\varepsilon_2\eta_1}{\alpha_2 g_3} +$
$+\dfrac{\varepsilon_3\eta_2\varepsilon_2\eta_1}{\alpha_3}]$

Here the 1$^{st}$, 2$^{nd}$ and 3$^{rd}$ terms describe the feedback of ions generated in the 1$^{st}$, 2$^{nd}$ and 3$^{rd}$ GEMs, respectively.

In case of identical holes in all GEMs we have

$\varepsilon_2 = \varepsilon_3 = \varepsilon$ ; $\eta_1 = \eta_2 = \eta$ ; $\alpha_1 = \alpha_2 = \alpha_3 = \alpha$ ;
$g_2 = g_3 = g$ ; $g_1 = g/b$ ; $g = (bG)^{1/3}$ ,

where parameter $b$ arises due to the electron collection efficiency of the 1$^{st}$ GEM different from that of other GEMs. Now (3) becomes

(4) $\quad F_{3GEM} = \dfrac{\varepsilon_1}{\alpha}[\dfrac{1}{g^2} + \dfrac{\varepsilon\eta}{g} + (\varepsilon\eta)^2] =$
$= a[(bG)^{-2/3} + c(bG)^{-1/3} + c^2] \quad .$

This expression can be checked in experiment. It has only 3 parameters, *a*, *b* and *c*, the latter characterizing the ion transfer efficiency of a single GEM. It is easy to show that expression (4) is valid not only for 3GEM+PCB structures, but for 3GEM structures as well. The ion feedback in 4GEM structures is described by a similar expression:

$$(5) \quad F_{4GEM} = a\,[(bG)^{-3/4} + c(bG)^{-2/4} + c^2(bG)^{-1/4} + c^3]$$

Fig.10 shows the fit of expressions (4) and (5) to experimental data for 3GEM+PCB, 3GEM and 4 GEM structures, with hole diameters of 85 and 70 μm. In these data sets, the electric fields in all gaps and inside the GEM holes were proportional to each other. Therefore in the first approximation the parameters can be considered to be independent of applied voltages. One can see that the gain dependence is well reproduced by the model, except of a few points at high gain. Note that the ion transfer efficiency of a single GEM, for given hole diameters and at given transfer fields, turned out to be rather large: $c$=0.3-0.5.

Disregarding last points (they will be discussed in the following section), one may conclude that the ion feedback fraction tends to a constant value at high gains. This value is the last term in (3)-(5), meaning that the ion feedback at high gains, in these particular GEM hole configurations, is induced mainly by the last GEM element.

In case of unequal hole diameters in different GEMs, expression (3) may be transformed to the following:

$$(6) \quad F_{3GEM} = aG^{-2/3} + bG^{-1/3} + c.$$

We cannot extract the charge transfer efficiency of a single GEM using this formula. We however can evaluate a contribution of each GEM to ion feedback. Fig.11 shows a comparison of the model to experimental data for GEM structures with reduced holes. One can see that the inverse power dependence on gain, revealed in the previous section, is well reproduced by the model. The matter is that the major contribution to ion feedback is provided now by the middle GEM element, giving rise to $G^{-1/3}$ dependence. This is because the real gain of the middle GEM, and consequently its ion current, is higher and its transparency for ions arrived from the 3$^{rd}$ GEM is lower as compared to other GEM elements. Note that the ion feedback current of the 1$^{st}$ GEM can be neglected (see the parameter values in Fig.11).

### 4.2. Avalanche extension

As we saw, starting from a certain critical gain, of about $5\times10^4$ in Ar/CF$_4$ and $2\times10^5$ Ar/CH$_4$, the ion feedback suppression is substantially enhanced: the last data points in Figs. 3, 5 and 10 are consistently lower than the general curve. We suppose that this effect might be connected to the avalanche extension outside the GEM holes, which also has the threshold in gain [3,13]. Indeed, if it would be the case, the positive ions produced outside the hole would have more chances to drift to the bottom GEM electrode rather than to enter the hole. It is interesting that the critical gain at which the supposed avalanche extension takes place here is of the same order as that estimated earlier in pure Ar using the pulse-shape analysis [13]: $4\times10^4$.

### 5. Conclusions

We have studied the feedback of positive ions in triple and quadruple GEM detectors. The dependence of ion feedback on some parameters was investigated in detail. These are the gas mixture, pressure, GEM hole diameter, detector gain, drift field, transfer field, 1$^{st}$ GEM voltage and number of GEMs. The principal observed effects and conclusions are listed below.

(1) The ion feedback is practically independent of the gas mixture and the pressure.

(2) The ion feedback is most sensitive to the drift field. It increases almost linearly with it.

Therefore mixtures providing operation at low drift fields, for example Ar/CH$_4$, should be used.

(3) The ion feedback decreases with gain. In a wide gain range, it can be described by an inverse power function of the gain. It is also rather sensitive to the GEM hole diameter. Other parameters, such as the transfer field and the number of GEMs, have minor effect.

(4) The configuration providing the highest suppression of ion feedback is a triple GEM structure with a reduced hole diameter of the middle GEM and reduced 1$^{st}$ GEM voltage. In this configuration the ion feedback, at a gain of 10$^4$, can be suppressed down to 0.5% at a drift field of 0.1 kV/cm and 3% at a drift field of 0.5 kV/cm. We think that this is the best that can be obtained in multi-GEM structures at the moment.

(5) A simple model of ion feedback in multi-GEM structures is proposed. The gain dependence of ion feedback is well reproduced. Using this model, it is possible to determine which GEM element provides the largest contribution to ion feedback and to estimate the ion transfer efficiency of a single GEM.

(6) At high gains, of the order of 10$^5$, the ion feedback suppression is enhanced. We suppose that this effect might be related to the avalanche extension from the GEM holes.

The results obtained are relevant to the performance of time projection chambers and gas photomultipliers, where the problem of ion feedback suppression is of primary importance. The level of ion feedback suppression achieved in the current work, of the order of 1%, may be enough for TPCs, but may not for gas photomultipliers with visible photocathodes. In the latter, the maximum gain and the life-time of the device is directly connected to the amount of ions hitting the photocathode. Therefore the ways towards reaching the ultimate suppression, *F~1/G*, should be paved.

One of the ways might be using a special shape of the GEM hole, namely a single conical shape with large input and small output apertures. Another possibility is coupling the multi-GEM structure to Micromegas [14]: the latter also has the natural capability for ion feedback suppression. Their combined suppression power could presumably be enhanced by at least an order of magnitude.

## Acknowledgements

This work has been done in the frame of R&D of the TESLA TPC.

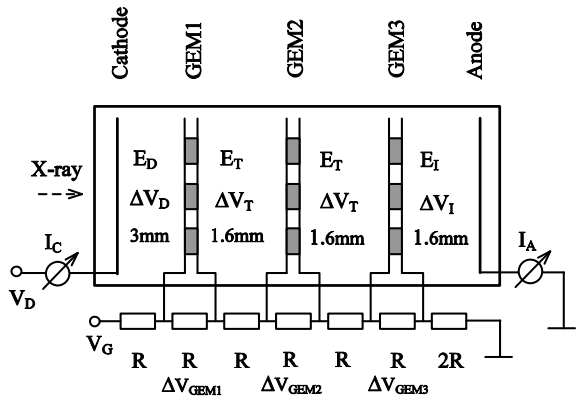

Fig.1 A schematic view of a triple GEM detector with the appropriate notations.

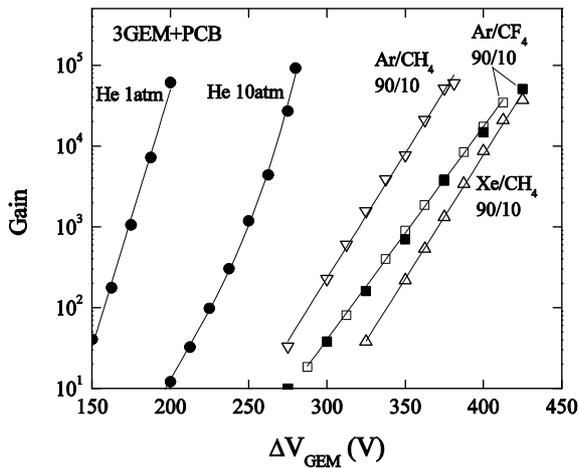

Fig.2 Gain of triple GEM detectors at different pressures in He [12] and in different gases at 1 atm as a function of the voltage across each GEM. Two GEM hole configurations were used: 85-85-85 μm (filled points) and 85-40-85 μm (open points).

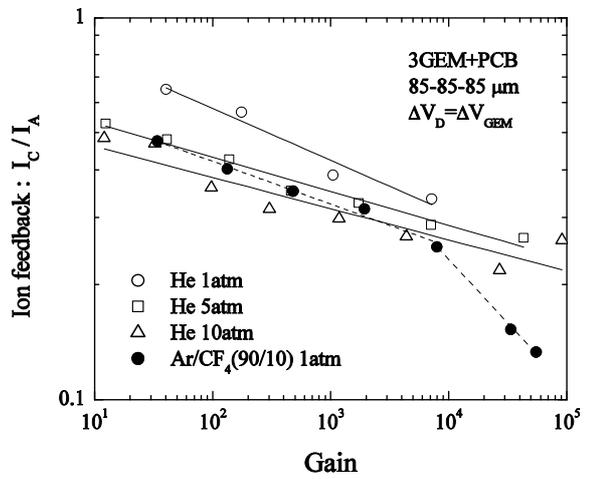

Fig.3 Effect of the pressure and the gas mixture. Ion feedback fraction as a function of the gain of a triple GEM detector at different pressures and in different gases, at $\Delta V_D = \Delta V_{GEM}$.

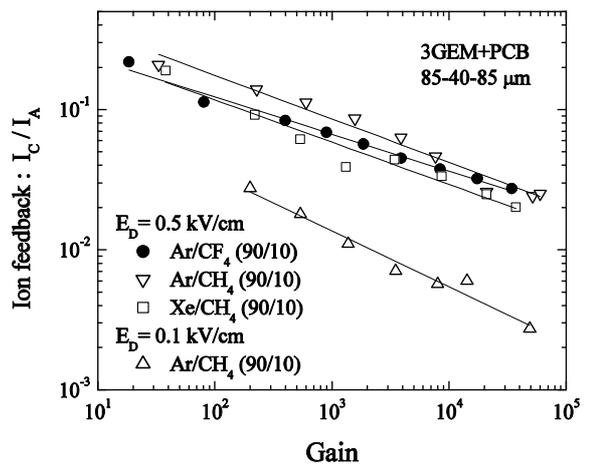

Fig.4 Effect of the gas mixture. Ion feedback fraction as a function of the gain of a triple GEM detector in different gases (at 1 atm), at a constant drift field $E_D$=0.5 kV/cm. In Ar/CH$_4$, the effect of reducing the drift field, down to 0.1 kV/cm, is demonstrated.

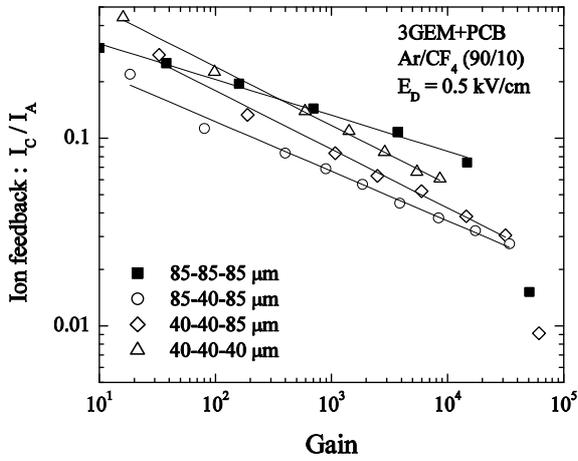

Fig.5 Effect of the GEM hole diameter. Ion feedback fraction as a function of the gain of a triple GEM detector at different GEM hole configurations.

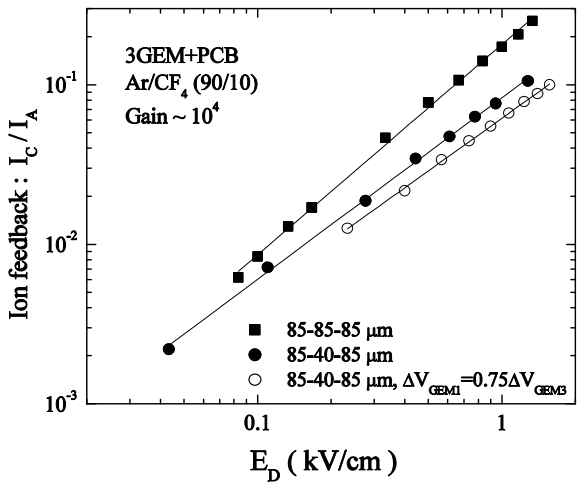

Fig.6 Effect of the drift field. Ion feedback fraction as a function of the drift field at a gain of $10^4$ of a triple GEM detector, at two GEM hole configurations. The configuration with decreased 1$^{st}$ GEM voltage is also shown.

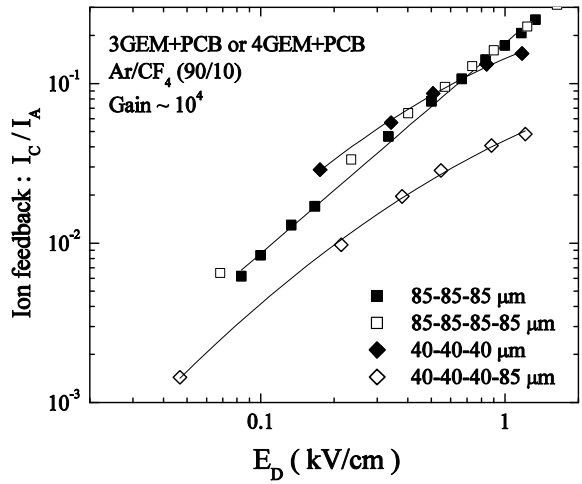

Fig.7 Effect of increasing the number of GEMs. Ion feedback fraction as a function of the drift field in triple and quadruple GEM detectors at a gain of $10^4$, at different GEM hole configurations.

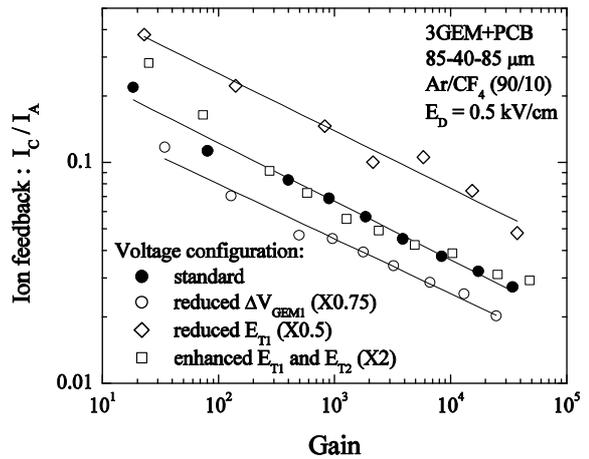

Fig.8 Effects of the transfer field and the 1$^{st}$ GEM voltage. Ion feedback fraction as a function of the gain of a triple GEM detector at different voltage configurations. Standard configuration corresponds to the voltage divider shown in Fig.1.

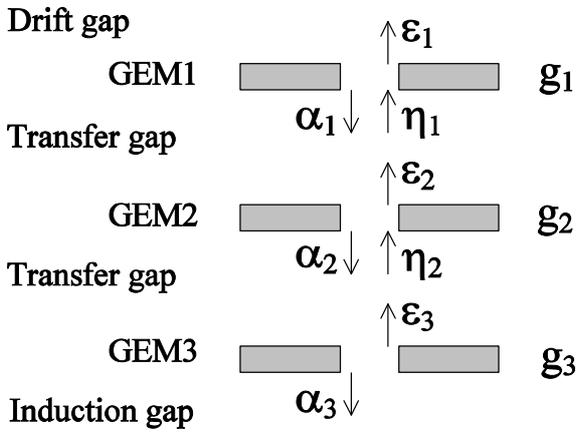

Fig.9 Notations used in the model of ion feedback.

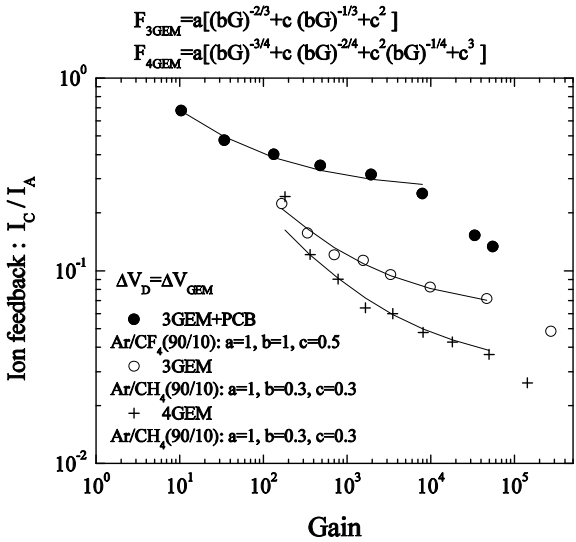

Fig.10 Comparison of ion feedback model (curves) to experimental data (points) for multi-GEM structures with identical holes: 3GEM+PCB (hole diameter 85 μm, current work); 3GEM and 4GEM (hole diameter 70 μm, [4]). Parameter $c$ is the charge transfer efficiency of a single GEM.

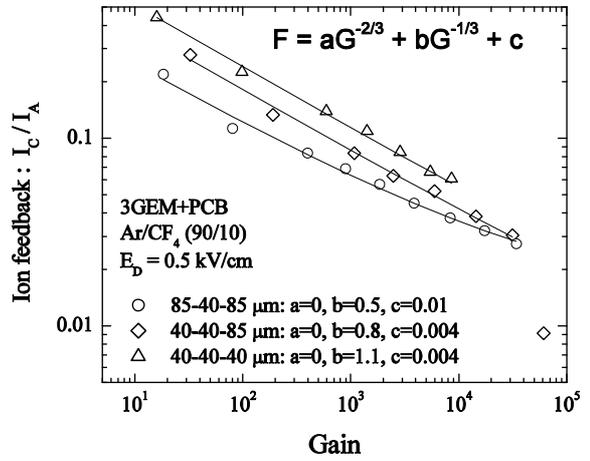

Fig.11 Comparison of ion feedback model (curves) to experimental data (points) for 3GEM+PCB structures with reduced holes.